\documentclass[]{svproc}

\usepackage{url}

\usepackage[utf8]{inputenc}   
\usepackage[T1]{fontenc}
\usepackage[pdftex]{graphicx}
\usepackage[ruled,vlined,linesnumbered]{algorithm2e}
\usepackage{booktabs}
\usepackage{makecell}
\usepackage{multirow}
\usepackage[hidelinks]{hyperref}
\usepackage[show]{chato-notes}
\usepackage{balance}
\usepackage{amsfonts}
\usepackage{enumerate}
\usepackage{cite}

\makeatletter
\newcommand{\removelatexerror}{\let\@latex@error\@gobble}
\makeatother

\newcommand{\suchthat}{\ensuremath{\,\mid\,}}

\begin{document}
\mainmatter 
\title{Decentralized and Self-adaptive Core Maintenance on Temporal Graphs}

\author{Davide Rucci\inst{1} \and Emanuele Carlini\inst{1} \and Patrizio Dazzi\inst{2} \and Hanna Kavalionak\inst{1} \and Matteo Mordacchini\inst{3}}

\authorrunning{Rucci, Carlini, Dazzi, Kavalionak and Mordacchini}

\institute{ISTI-CNR, Pisa, Italy\\ 
\email{davide.rucci@isti.cnr.it},
\email{emanuele.carlini@isti.cnr.it},
\email{hanna.kavalionak@isti.cnr.it}
\and
University of Pisa, Pisa, Italy\\
\email{patrizio.dazzi@unipi.it}
\and
IIT-CNR, Pisa, Italy\\
\email{matteo.mordacchini@iit.cnr.it}
}

\maketitle

\begin{abstract}
%
Key graph-based problems play a central role in understanding network topology and uncovering patterns of similarity in homogeneous and temporal data. Such patterns can be revealed by analyzing communities formed by nodes, which in turn can be effectively modeled through temporal $k$-cores. 
This paper introduces a novel decentralized and incremental algorithm for computing the core decomposition of temporal networks.
Decentralized solutions leverage the ability of network nodes to communicate and coordinate locally, addressing complex problems in a scalable, adaptive, and timely manner.
By leveraging previously computed coreness values, our approach significantly reduces the activation of nodes and the volume of message exchanges when the network changes over time. This enables scalability with only a minimal trade-off in precision. Experimental evaluations on large real-world networks under varying levels of dynamism demonstrate the efficiency of our solution compared to a state-of-the-art approach, particularly in terms of active nodes, communication overhead, and convergence speed.

%
%
%
%
%

\end{abstract}


\section{Introduction}\label{sec:introduction}

Graphs are a fundamental data model in computer science, capable of representing virtually any relationship among a set of entities. Their structure allows many key problems to be addressed in a graph-based form, such as detecting connected components~\cite{ugurlu2025distributed, cracker_tpds}, computing node centrality~\cite{audrito2021aggregate}, solving vertex cover problems~\cite{yigit2021breadth,carlini2024efficient}, and performing {\em core decomposition}~\cite{SEIDMAN1983269,montresor_distributed_kcore_2013,batagelj2003omalgorithmcoresdecomposition}.
These problems are closely related because their focus is on understanding the underlying organization of the graph, particularly in identifying communities—subsets of nodes that are densely connected internally but sparsely connected to the rest of the network.
The detection and analysis of such community structures are crucial in many real-world applications, as proven by the many research efforts in the field.
However, as graph data becomes increasingly large and available, traditional centralized algorithms face significant limitations in terms of scalability, latency, and resource consumption. In response, decentralized computing models have emerged as a viable alternative. In fact, by distributing the computational workload across the network, decentralized systems enhance scalability and enable more efficient processing of large-scale graphs~\cite{xu2024decentralized,vincent2023systematic}.

A further layer of complexity arises when considering dynamic or temporal graphs, where nodes and edges change over time. This temporal dimension introduces new challenges for graph analysis, particularly for centralized systems that rely on global knowledge and batch processing. Centralized methods often struggle to accommodate rapid or frequent changes, resulting in inefficiencies and delayed responses. In contrast, decentralized approaches can naturally adapt to dynamic environments by leveraging localized communication and incremental updates. This adaptability allows nodes to recompute relevant metrics in response to changes without restarting from scratch, thereby maintaining efficiency in continuously evolving networks.

Within this context, this paper focuses on the decentralized \emph{coreness} (or core number) computation and its evolution over time. 
Since their introduction in the 1970s~\cite{SEIDMAN1983269}, $k$-cores and their analysis have gained increasing popularity. Knowing a node’s coreness -- i.e., the largest $k$ such that it belongs to the $k$-core -- offers valuable insight into the network structure, supporting tasks like community detection and node ranking~\cite{SantoFortunato,KONG20191,giatsidis_evaluating,k_core_survey2020}. Coreness reflects how densely connected a node is within its local neighborhood and indicates membership in cohesive subgraphs, making it a useful indicator of structural relevance.

In social networks, the coreness helps identify influential users that facilitate information spread~\cite{SantoFortunato}; it is also applied, among other uses, to anomaly detection~\cite{faloutsos_corescope} and routing optimization in blockchain networks~\cite{state_sharding_blockchain}.
Tracking coreness over time further aids in understanding the evolution of core structures and identifying stable, central components~\cite{Rucci:2024SAC}. While some decentralized algorithms have been proposed for static $k$-core computation (e.g., Montresor et al.~\cite{montresor_distributed_kcore_2013}), they do not handle temporal changes natively. Conversely, several centralized approaches exist for temporal $k$-core decomposition~\cite{bonchi_span_core}, but they lack the scalability of decentralized methods.

We introduce a decentralized, iterative, message-passing, and incremental algorithm for computing the $k$-core composition of temporal graphs. Our approach leverages previously computed coreness values to reduce the number of active nodes and exchanged messages, accepting a slight trade-off in accuracy. 
In fact, our experiments on large, real-world graphs show a reduction of 50\%-90\% in the number of total messages exchanged during the execution of our algorithm with respect to a state-of-the-art competitor.
%

To summarize, the main contributions of this paper are: 
\begin{itemize} 
    \item A novel decentralized algorithm for coreness computation in temporal graphs; 
    \item Simulation over large, heterogeneous real-world networks with varying temporal parameters; 
    \item A comparative evaluation with a state-of-the-art approach, analyzing active nodes, message counts, and convergence; 
    \item Public release of the code to foster reproducibility and comparison. 
\end{itemize}

The remainder of this paper is structured as follows. Section 2 reviews the related work. Section 3 outlines the algorithm, notation, and definitions. Section 4 details the experimental setup and presents the main results. Finally, Section 5 concludes with a summary of key contributions and future research directions.

\section{Related Work}\label{sec:related_work}

Several studies have explored core decomposition in temporal networks, developing various definitions and centralized algorithms \cite{malliaros2020core}.
Yang et al.~\cite{yang2023scalable} introduce the Temporal $k$-core Query problem and propose the Temporal Core Decomposition algorithm, which efficiently computes $k$-cores across time intervals while minimizing redundant calculations. Li et al.~\cite{li2018core_union} define a $k$-core model that ensures stability over time, using graph reduction techniques coupled with a branch-and-bound approach. Wu et al.~\cite{wu2015core} present a temporal core decomposition method that incorporates edge-based constraints to identify frequently interacting subgraphs. 

Galimberti et al.~\cite{bonchi_span_core} propose a method for analyzing temporal networks using temporal core decomposition.
In their framework, each core is characterized by two parameters: the minimum degree and the span, which indicates the time interval over which the core persists. They develop efficient algorithms to compute these \emph{span-cores} and demonstrate their effectiveness in real-world applications, such as analyzing contact networks and studying the evolution of social interactions.
Although the approach of Galimberti et al. aligns with our philosophy of intersection-based decomposition (see Section~\ref{subsec:notation}), their definition of temporal core identification does not match our definition, since they have tailored it to compute a slightly different concept.
A recent work by Conte et al.~\cite{Rucci:2024SAC} offers a comprehensive summary of the various definitions of temporal $k$ core that have been proposed in the above articles, highlighting the key differences and comparing the results that can be obtained by exploiting those definitions. 
This work also drives our algorithm design later in Section~\ref{sec:algorithm}.

These contributions mostly focus on centralized algorithms for the temporal core decomposition task, assuming that a single computational entity processes the entire graph.
Our work shifts the focus to decentralized solutions, aiming to distribute the computation across multiple nodes rather than relying on a central processing unit.

An influential solution to decentralize the computation of the $k$-core for static graphs was introduced by Montresor et al. \cite{montresor_distributed_kcore_2013}. 
They proposed a message-exchange algorithm based on the locality property of the $k$-core decomposition, which states that the coreness of a node is the highest number $k$ for which the node has at least $k$ neighbors in a $k$-core or higher. 
%
While their algorithm offers a fast convergence rate, it cannot be directly translated into a temporal scenario, as it would require recomputing everything from scratch each time the graph changes.

%

Aridhi et al. \cite{10.1145/2933267.2933299} propose distributed algorithms for efficient core decomposition and maintenance in large-scale dynamic graphs. Their approach addresses the computational challenges of massive and evolving graph structures by incrementally updating the core decomposition as the graph changes. However, such a solution relies on a hybrid model with a master node that orchestrates the update process, while different worker nodes handle the partitions. In contrast, our work proposes a fully decentralized solution, focusing on reducing the number of activated nodes and minimizing message exchanges, further optimizing performance in dynamic graph settings.
%
Weng et al.~\cite{Weng:TPDS2021} leverage a Pregel-like graph computing framework and focus on efficient algorithms to update core decompositions as the graph evolves. Their proposed methods aim to enhance the performance and scalability of core maintenance tasks within distributed computing environments, addressing challenges in large-scale graph processing.
%
Liu and Zhang~\cite{Liu:2020IEEEAccess} introduce core maintenance algorithms for edge insertion and deletion, which are validated through experiments on real-world graphs, in dynamic, edge-weighted graphs. 
This is different from our approach, which is focused on unweighted temporal graphs.
%
Yu et al.~\cite{Yu:2022TCSS} present efficient algorithms for maintaining core numbers in dynamic graphs. They introduce the concept of a superior edge set to handle multiple edge insertions and deletions simultaneously, reducing redundant vertex visits. 
Their incremental and decremental core maintenance algorithms support parallel implementations, but remain centralized and are primarily designed for dynamic graphs, unlike our approach, which emphasizes decentralized solutions on temporal graphs.


In summary, while these studies offer valuable contributions to the core decomposition and maintenance task in dynamic and temporal networks, they predominantly rely on centralized or semi-centralized methods. In contrast, our approach is decentralized, focusing on minimizing the number of activated nodes and reducing redundant message exchanges. By distributing the computational load across multiple nodes and optimizing communication, we aim to achieve greater efficiency and scalability in dynamic environments, without depending on a central processing entity.
Another key distinction lies in the concepts of temporal and dynamic graphs. Temporal graphs are inherently dynamic, but organized differently.
Firstly, temporal graphs typically allow for the reconstruction of changes over time by providing snapshot- or epoch-based organization.
On the other hand, dynamic graphs are usually designed to react to instant changes, also providing algorithmic methods to update necessary data structures whenever a single edge or node is inserted or deleted.
Furthermore, in temporal graph algorithms, we can decide how significant an interaction is within a specific time interval. For example, we can prioritize a connection that lasts longer over many connections that exist only for a single snapshot.

\section{Algorithm}\label{sec:algorithm}
This Section outlines our decentralized approach for determining the coreness of nodes in a temporal graph. We begin with fundamental notation and definitions that will be referenced throughout the paper, followed by a presentation and analysis of our algorithm's pseudocode.

\begin{algorithm}[!ht]
    \caption{Distributed Algorithm for $k$-core computation in a temporal graph $G_\tau$, run by each node $u \in V$.}\label{alg:rucci}
    \DontPrintSemicolon
    \SetKwProg{MyFn}{Function}{}{end}
    \SetKwFunction{ComputeCoreness}{computeCoreness}
    \SetKwFor{On}{on}{do}{}
    \SetKw{Send}{send}
    \SetKwBlock{Rep}{repeat}

    \On(\tcp*[h]{when a node joins $G_\tau$}){initialization}{
        changed $\gets$ false\;
        core $\gets d(u)$\;
        \lForEach{$v \in N(u)$}{est$[v] \gets \infty$}
    }

    \On{epoch change}{
        oldCore $\gets$ core\;
        \If{there is at least one new neighbor $v$}{\label{alg:rucci:line:new-neighbor}
            core $\gets d(u)$\; 
            changed $\gets$ true\;
            est$[v] \gets \infty$\;
        }
        \ElseIf{a neighbor is lost}{\label{alg:rucci:line:lost-neighbor}
            core $\gets$ \ComputeCoreness(est,u,core)\;
            \lIf{oldCore $\neq$ core}{changed $\gets$ true}
        }
    }

    \On{receive $\langle v, k\rangle$}{
        est$[v] \gets k$\;
        $t \gets$ \ComputeCoreness(est,u,core) \;  
        \If{$t \neq core$}{
            \lIf{$t < core$}{wait one iteration before sending update}\label{alg:rucci:line:wait}
            core $\gets t$\;
            changed $\gets$ true\;
        }
    }

    \Repeat{convergence}{\label{alg:rucci:line:repeat}
        \If{changed}{
            \Send{} $\langle u, core \rangle$ \KwTo $N(u)$\;
            changed $\gets$ false\;
        }
    }

    \MyFn{ComputeCoreness(est, u, k)}{\label{alg:rucci:line:computecoreness}%
        \ForEach{$e \in est[]$}{
            \If{$e < \infty$}{
                \eIf{$e < d(u)$}{
                    count$[e] \gets $ count$[e] + 1$
                }
                {
                    count$[d(u)] \gets$ count$[d(u)] +1$
                }
            }
            \lElse{
                \KwRet $\min \{core, d(u)\}$
            }
        }

        total $\gets 0$\;
        \For{$i = $ count.length \textbf{down} \KwTo $0$}{
            total $\gets $ total $+ $ count$[i]$\;
            \lIf{total $\geq i$}{\KwRet i}
        }
        
        \KwRet $d(u)$ 
    }

\end{algorithm}

\subsection{Notation and Definitions}\label{subsec:notation}
Formally, we define a temporal graph $G_\tau$ as a pair of sets of vertices and temporal edges $(V, E_\tau)$, where $\tau \in \mathbb{N}$ is called the \emph{lifespan} of $G$, and the set of edges of $G_\tau$ is defined as $E_\tau = \left\{ (u,v,t) \suchthat u,v \in V, 1 \leq t \leq \tau \right\}$, where $t$ is the \emph{timestamp} of an edge.
If we consider only the edges of $E_\tau$ with a fixed value for $1 \leq t \leq \tau $, we obtain the static graph $G_t = (V, E_t)$ that is called \emph{snapshot} or \emph{epoch} of $G$ at time $t$.
A static \emph{$k$-core} $K$ is defined as the inclusion-maximal subset of vertices $K \subseteq V$ such that every vertex of $K$ has degree at least $k$ in the \emph{vertex-induced subgraph} $G[K] = (K, E_K = \left\{ (u,v) \in E \suchthat u,v \in K\right\})$.
The maximum $k$ for which a node $v$ belongs to the $k$-core of a graph is called \emph{coreness} (also \emph{core number} in the literature~\cite{SEIDMAN1983269}) of $v$.
The main issue to be addressed when translating the concept of $k$-cores from static to temporal graphs is the selection of the most appropriate aggregation function, i.e., which edges are to be considered for the computation of cores for a given time interval.
A recent study by Conte et al. \cite{Rucci:2024SAC}, empirically shows that there exists no one-for-all solution to this problem; instead, a set of graphs may be analyzed with a set of distinct \emph{aggregation functions} $af: E_\tau \times [a, b] \rightarrow E_\tau$ that, given a time interval $[a,b]$ tell us which edges to consider as present in the graph for that interval.
We informally summarize here the possible functions that can be adopted, referring the reader to the paper by Conte et al.~\cite{Rucci:2024SAC} for a more formal overview of those and when to use which.
Given a time interval $[a,b], 1\leq a \leq b \leq \tau $, we can consider the edge $(u,v,t) \in E_\tau$ to exist in that interval if:
\begin{itemize}
    \item\emph{(intersection)}: the edge $(u,v,t) \in E_\tau$ for \emph{all} $t = a,\dots,b$; 
    \item \emph{(union)}: the edge $(u,v,t) \in E_\tau$ for \emph{any} $t=a,\dots,b$;
    \item \emph{(union-$h$)}: the edge $(u,v,t) \in E_\tau$ for \emph{at least} $h$ distinct values of $t=a,\dots,b$. We will denote this function by $\cup_h$ later in the paper.
\end{itemize}
We denote the \emph{neighborhood at time $t$} of $v \in V$ as $N_t(v)$, or just $N(v)$ whenever $t$ is clear from the context. 
Similarly, the \emph{degree at time $t$} of $v \in V$ is $d_t(v) = |N_t(v)|$, omitting the $t$ when clear from the context.
A key question to address each time we use intervals on temporal graphs is: how do we choose both the extremes and the length of intervals?
Our answer is to introduce a parameter, called \emph{memory size}, to serve as the length of every interval; then, we use a sliding window approach, spanning the entire lifetime of the graph. 
For example, setting the memory size to $5$ implicitly allows nodes to ``recall'' the last 5 snapshots in their neighborhood. Then, we use the edge aggregation function to decide which neighbors are actually part of the graph in the interval.

\subsection{Our Decentralized Algorithm}

Our strategy is formalized in the pseudocode of Algorithm~\ref{alg:rucci}.
%
%
It is a message-exchange approach in which the computation is organized in rounds or \emph{iterations}. For each iteration, the nodes communicate estimates of their coreness value to their neighbors, which, in turn, adjust theirs based on what they received.
We note that Algorithm~\ref{alg:rucci} is fully decentralized and can run for an indefinite period of time, as it can adapt to every change in the graph.
There are three main parts into which we can subdivide the algorithm that runs from every node present in the graph: \emph{initialization}, \emph{new epoch}, and \emph{reception of a message}, and we analyze them separately.
\paragraph{Initialization.} When a new node entity is generated, the initial step is performed: it assigns its coreness estimate to its degree, as this is the sole information available at creation, sets its neighbors' estimates to infinity since they are unknown, and then communicates its degree to its neighbors.
\paragraph{Epoch Change.} When the event of a new epoch in the temporal graph occurs, there are three possibilities for each node: \emph{(a)} the node gains a new neighbor, \emph{(b)} the node does not gain any new neighbor but loses at least one neighbor instead, or \emph{(c)} the neighborhood of the node is not affected. Only the first two possibilities trigger a reaction of the node:
\begin{enumerate}[(a)]
    \item the node acquires at least one new neighbor, prompting it to update its coreness estimate to reflect its new degree, as this ensures accuracy.
    \item the node experiences a loss of one or more neighbors without gaining any new ones; under these circumstances, the node attempts to recalculate its coreness using knowledge of its remaining neighbors. This capability is crucial to the algorithm, allowing us to use preexisting graph information and bypass the need to transmit potentially unnecessary messages. This approach also helps in identifying whether the departure of neighbors affects coreness, without the need to wait for a complete message cycle.
\end{enumerate}
Note that case \emph{(a)} takes precedence over case \emph{(b)}, since the simultaneous occurrence of these events results in at least one infinite value in the node's estimate table, potentially raising its coreness value. Therefore, we opt for a cautious strategy by resetting the coreness to the degree.
\paragraph{Message received.} The last event that we consider is the reception of a message from a neighbor, which triggers an update in the table of estimates of the recipient and possibly a change in the coreness value. 
The main aspect is the check on line~\ref{alg:rucci:line:wait}, which postpones sending an update message to the following iteration if our coreness estimate is less than before.
This helps reduce potential errors originating from delayed change propagation across the graph, since a neighbor might experience an increase in its coreness when a new node joins its neighborhood.
Indeed, it requires two iterations for a node to detect a change in the coreness of its neighbors if those neighbors have acquired new connections within the same epoch. We will address this issue later.

\subsection{Example}\label{subsec:example}

\begin{figure}[tb]
    \centering
    \includegraphics[width=0.99\textwidth]{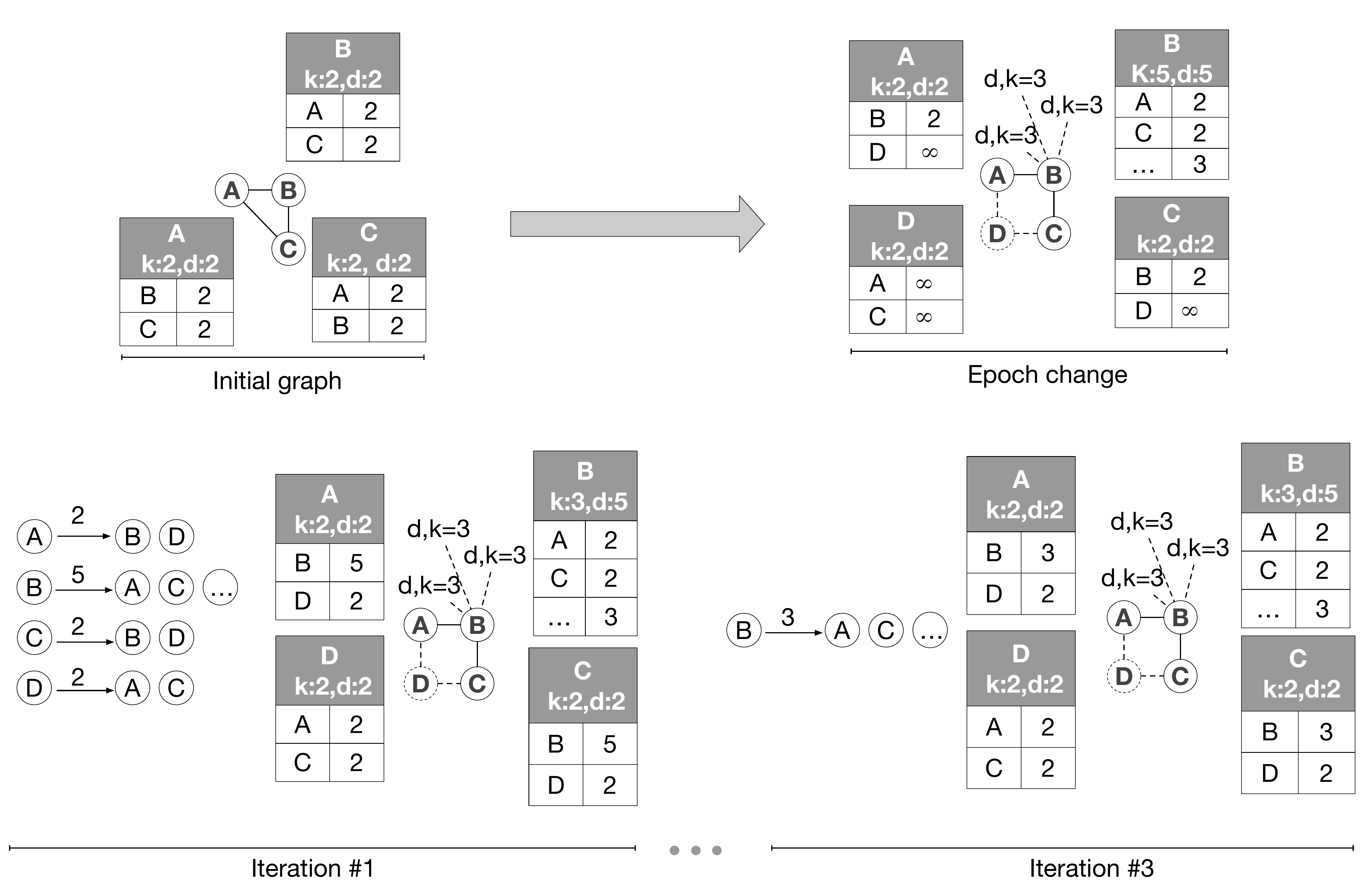}
    \caption{An example of Algorithm~\ref{alg:rucci} in action. Dashed nodes and edges are inserted in the new epoch.}
    \label{fig:example}
\end{figure}

Fig.~\ref{fig:example} illustrates our algorithm in action. The initial graph consists of three nodes, \textbf{A}, \textbf{B}, and \textbf{C}, all interconnected.
In the new epoch, a new node \textbf{D} arrives, connecting with nodes \textbf{A} and \textbf{C}, while the link between \textbf{A} and \textbf{C} disappears. Furthermore, three new neighbors of \textbf{B} appear, each with a degree and coreness of 3. For the sake of clarity and simplicity, in this example we ignore the updates sent by the new neighbors of \textbf{B}, and show only the local effects on the initial nodes.
Since each node has acquired at least one new neighbor, they all revert their coreness to match their degree and adjust the table entry for any new neighbors to infinity.
In iteration \# 1, all nodes that received a new neighbor (and thus have their \texttt{changed} variable now set to \texttt{true}) send a new message with their core estimate (set to their degree) to their neighbors. Nodes \textbf{A}, \textbf{C}, and \textbf{D} send a value of 2, while \textbf{B} sends a value of 5.
As a result, all nodes get messages and revise their coreness. Nodes \textbf{A}, \textbf{C}, and \textbf{D} have unchanged coreness and therefore cease sending messages.
Node \textbf{B} alters its coreness to 3, down from 5. Consequently, it delays messaging its neighbors for one iteration to account for possible propagation delays due to changes occurring in its two-hop neighbors.
The message is finally sent at Iteration \#3. Since this new information received does not change the coreness of any neighbors, the process stops.

\subsection{Efficiency/errors trade off}
Since Algorithm~\ref{alg:rucci} is the result of empirical and iterative refinements, its strategy may lead to errors in the computed coreness for some nodes.
This primarily results from each graph node consistently treating the estimates it receives from its neighbors as valid, rather than resetting them to infinity with each epoch change.
Thus, while executing the function on line~\ref{alg:rucci:line:computecoreness}, a node's coreness may abruptly shift from a high value, such as its degree, to a minimal coreness value, bypassing potential intermediate values.
Generally, this isn't a problem since the coreness can vary; however, this sometimes causes nodes to compute incorrect values.
Our empirical analysis revealed that this phenomenon typically occurs when new nodes enter the graph. Neighbors, located two hops from these new nodes, tend to calculate an incorrect value compared based on the competitor we considered, which also serves as the ground truth for coreness values.
This explains why we implemented a delay before dispatching updates when a node calculates a coreness lower than its prior value: this allows the node to potentially determine a new and accurate coreness in the next cycle if it gets additional updates from its neighbors.
Adopting a method that allows for some errors involves a compromise between strategy efficiency and error count: indeed, to completely eliminate errors, we would need to reset all estimates whenever an epoch changes, requiring each node to send at least one message for every epoch in the graph.
Conversely, by extensively reusing our existing estimates, we risk calculating an incorrect value, but gain a notably faster algorithm.

As we verified (see Section~\ref{sec:experiments}) that the offset of the errors produced by Algorithm~\ref{alg:rucci} is always in the range $\pm 1$ and that, most importantly, it affects a very small portion of nodes in the whole graph, we accept to have some small errors to get a big saving in the efficiency metrics in return.
\begin{table}[hbt]
    \caption{Overview of our dataset.}\label{tab:dataset}
    \centering
    \resizebox{\columnwidth}{!}{%
    \begin{tabular}{llrr}
        \toprule
            Dataset & Kind & Max. Nodes & Total Edges \\
        \midrule
            AS-733 & Autonomous System & 7716 & 11410810\\
            Email-EU-Core & Email Exchange & 986 & 332334 \\
            Rec-Amazon-Ratings & Amazon Product Reviews & 2146057 & 5838027 \\
            sx-mathoverflow & Q\&A & 24759 & 390441  \\
            reddit & Hyperlinks between subreddits & 54075 & 858488 \\ 
        \bottomrule
    \end{tabular}%
    }
\end{table}

\section{Experimental evaluation}\label{sec:experiments}
This section provides an overview of our comprehensive experimental phase, describes the algorithm selected as the competitor for validating our analysis, and presents a discussion of the key results achieved.

\subsection{Competitor}
To compare and verify our results, we take advantage of a revised version of the algorithm originally proposed by Montresor et al.~\cite{montresor_distributed_kcore_2013}, which was designed for static graphs.
The structure of this algorithm mirrors that of Algorithm~\ref{alg:rucci}, meaning each node monitors updates to its own coreness estimate and maintains an estimate of the coreness of its neighboring nodes.
Montresor et al. provide a formal proof demonstrating that their method consistently converges to the accurate coreness value for every node in a graph \cite{montresor_distributed_kcore_2013}.
This algorithm can be modified to accommodate the temporal scenario by recalculating completely whenever there is an epoch transition: each node discards all estimates about its neighbors and assigns its coreness equal to its degree.
We use this adaptation to compute ground truth values for the coreness of nodes, and to show the amount of savings we can achieve by adopting our fine-tuned strategy instead of a direct translation of an existing algorithm to the temporal context.

\begin{table}[htb]
\centering
\caption{Summary of results obtained in our experiments. $\cup_h$ is the union-$h$ function (see Section~\ref{subsec:notation}). Results are averaged over all the executions of algorithms for all epochs. The memory size parameter is set to $5$. Error percentage is relative to the number of nodes in the graph.}
\label{tab:experimental_results}
\resizebox{\textwidth}{!}{%
\begin{tabular}{lcrrrrrrrrrrr}
\hline
\multicolumn{1}{c}{\multirow{2}{*}{Dataset}} &
  \multirow{2}{*}{\begin{tabular}[c]{@{}c@{}}Aggr.\\ Function\end{tabular}} &
  \multicolumn{1}{c}{\multirow{2}{*}{Epochs (length)}} &
  \multicolumn{3}{c}{Avg. Activated Nodes} &
  \multicolumn{3}{c}{Avg. Total Messages} &
  \multicolumn{3}{c}{Avg. Iterations} &
  \multicolumn{1}{c}{\multirow{2}{*}{\begin{tabular}[c]{@{}c@{}}Avg. Errors\\ per Epoch\end{tabular}}} \\ \cline{4-12}
\multicolumn{1}{c}{} &
   &
  \multicolumn{1}{c}{} &
  \multicolumn{1}{c}{Alg. 1} &
  \multicolumn{1}{c}{Comp.} &
  \multicolumn{1}{c}{Ratio} &
  \multicolumn{1}{c}{Alg.1} &
  \multicolumn{1}{c}{Comp.} &
  \multicolumn{1}{c}{Ratio} &
  \multicolumn{1}{c}{Alg.1} &
  \multicolumn{1}{c}{Comp.} &
  \multicolumn{1}{c}{Ratio} &
  \multicolumn{1}{c}{} \\ \hline
\multicolumn{1}{l|}{AS-733} &
  \multicolumn{1}{c|}{$\cap$} &
  \multicolumn{1}{r|}{105 (7 days)} &
  372 &
  3968 &
  \multicolumn{1}{r|}{0.09} &
  582 &
  4950 &
  \multicolumn{1}{r|}{0.11} &
  10 &
  7 &
  \multicolumn{1}{r|}{1.39} &
  13.7 (0.2\%) \\
\multicolumn{1}{l|}{AS-733} &
  \multicolumn{1}{c|}{$\cup$} &
  \multicolumn{1}{r|}{105 (7 days)} &
  431 &
  4630 &
  \multicolumn{1}{r|}{0.09} &
  741 &
  5912 &
  \multicolumn{1}{r|}{0.12} &
  14.1 &
  9.35 &
  \multicolumn{1}{r|}{1.51} &
  16.6 (0.3\%) \\
\multicolumn{1}{l|}{email-Eu-core} &
  \multicolumn{1}{c|}{$\cap$} &
  \multicolumn{1}{r|}{75 (7 days)} &
  91 &
  249 &
  \multicolumn{1}{r|}{0.38} &
  135 &
  331 &
  \multicolumn{1}{r|}{0.40} &
  7.5 &
  5.5 &
  \multicolumn{1}{r|}{1.37} &
  4.21 (0.4\%) \\
\multicolumn{1}{l|}{sx-mathoverflow} &
  \multicolumn{1}{c|}{$\cup$} &
  \multicolumn{1}{r|}{334 (7 days)} &
  701 &
  1481 &
  \multicolumn{1}{r|}{0.47} &
  1401 &
  2530 &
  \multicolumn{1}{r|}{0.55} &
  24.8 &
  13.8 &
  \multicolumn{1}{r|}{1.79} &
  9.92 (0.1\%) \\
\multicolumn{1}{l|}{reddit} &
  \multicolumn{1}{c|}{$\cup_2$} &
  \multicolumn{1}{r|}{174 (7 days)} &
  813 &
  2103 &
  \multicolumn{1}{r|}{0.38} &
  1291 &
  2799 &
  \multicolumn{1}{r|}{0.46} &
  15.5 &
  8.5 &
  \multicolumn{1}{r|}{1.81} &
  10.04 (0.02\%) \\
\multicolumn{1}{l|}{reddit} &
  \multicolumn{1}{c|}{$\cup_2$} &
  \multicolumn{1}{r|}{87 (14 days)} &
  813 &
  2103 &
  \multicolumn{1}{r|}{0.38} &
  1291 &
  2799 &
  \multicolumn{1}{r|}{0.46} &
  15.5 &
  8.5 &
  \multicolumn{1}{r|}{1.81} &
  10.1 (0.02\%) \\
\multicolumn{1}{l|}{reddit} &
  \multicolumn{1}{c|}{$\cap$} &
  \multicolumn{1}{r|}{174 (7 days)} &
  62 &
  186 &
  \multicolumn{1}{r|}{0.33} &
  82 &
  219 &
  \multicolumn{1}{r|}{0.38} &
  6.15 &
  4 &
  \multicolumn{1}{r|}{1.52} &
  0.5 (0.001\%) \\
\multicolumn{1}{l|}{rec-amazon-ratings} &
  \multicolumn{1}{c|}{$\cup$} &
  \multicolumn{1}{r|}{124 (28 days)} &
  69029 &
  220240 &
  \multicolumn{1}{r|}{0.31} &
  105124 &
  285357 &
  \multicolumn{1}{r|}{0.36} &
  45.9 &
  24.7 &
  \multicolumn{1}{r|}{1.85} &
  1344.5 (0.1\%) \\ \hline
\end{tabular}%
}
\end{table}

\subsection{Experimental Setup and Dataset}
We implemented both algorithms (Algorithm~\ref{alg:rucci} and the aforementioned competitor by Montresor et al.~\cite{montresor_distributed_kcore_2013}) in the Rust programming language, and carried out our experiments on the following architecture: Intel(R) Core(TM) i9-9900K CPU @ 3.60GHz, 8 physical cores, 16 logical cores with 64~GB of RAM, and 16~MB of shared L3 cache. Our code is publicly available on GitHub\footnote{\url{https://github.com/DavideR95/temporal_distributed_kcores}}.

To simulate the distributed environment, we used two queues for gathering nodes and messages sent by the nodes, to dispatch them to the appropriate recipients. 
Initially, we gather the indices of nodes required to transmit a message as a vector. Subsequently, each node sends its message to a separate message queue, and ultimately, each message is sent to the appropriate recipient for processing.
A node that receives a message and needs to send a new one will postpone its transmission until the next iteration of the algorithm.
To streamline deployment and evaluation, our implementation includes a central entity called the \emph{Graph} that verifies the convergence of the algorithm and the coreness computation of each node. However, we remark that Algorithm~\ref{alg:rucci} is entirely decentralized and works correctly without an orchestrator.

We run the experiments on a dataset of real-world temporal graphs obtained from the SNAP Repository~\cite{snapnets} and Network Repository \cite{networkrepository}.
%
%
The dataset, summarized in in Table~\ref{tab:dataset}, offers a varying level of dynamism, i.e. how much and how fast a graph changes over time, that can be noticed by looking at the plots in Figs.~\ref{fig:as733-comparison}(d) and~\ref{fig:mathoverflow-comparison}(d), which show how much a graph changes between two consecutive epochs.
For each graph, we only give the maximum number of distinct nodes that have been active at least once throughout the lifespan of the graph.
For any given epoch, the number of nodes that are part of the graph, i.e., that have at least one incident edge, is upper-bounded by this value.
The number of edges reported follows a similar rationale, representing the total count of edges present in the graph during at least one epoch.
We manually chose the epoch length for each graph by grouping edges according to their timestamp, in a way such that each epoch is sufficiently populated with nodes and edges so that the results obtained are not trivial (e.g. nodes are not isolated and there is enough change in active edges for any two consecutive epochs).
%
We run both algorithms on the whole dataset, collecting the following metrics:
\begin{itemize}
    \item\emph{Activated nodes}: nodes that sent at least one message during the execution of the algorithms.
    \item\emph{Number of iterations}: number of iterations of the main loop of the algorithms needed at a certain epoch to terminate their execution.
    \item\emph{Number of messages}: the number of messages sent by all nodes for each iteration of the algorithms.
    \item\emph{Errors}: number of nodes that computed a different coreness value in Algorithm~\ref{alg:rucci} with respect to the ones computed by our competitor.
\end{itemize}
Different values for the \emph{memory size} of each node were tested, spanning from a minimal size (1) to a larger size (10).
Given space constraints, we present results solely for a memory size of $5$, which represents a compromise between the extremes.
Moreover, as we pointed out in Section~\ref{sec:introduction}, we tested different edge aggregation functions, namely intersection, union and union-$2$ (or \emph{half}), i.e., every edge must appear at least half of the memory size times (for a memory size of 5, \emph{half} corresponds to 2, $= \lfloor 5/2 \rfloor$).

\subsection{Results and Discussion}
Table~\ref{tab:experimental_results} summarizes the results we obtained for Algorithm~\ref{alg:rucci}, compared to our competitor, highlighting the ratio between the two.
We can immediately notice that the ratio of both the activated nodes and the total messages is always well below 1, showing a big savings on these two critical measures.

\begin{figure}[!ht]
    \centering
    \includegraphics[width=\textwidth]{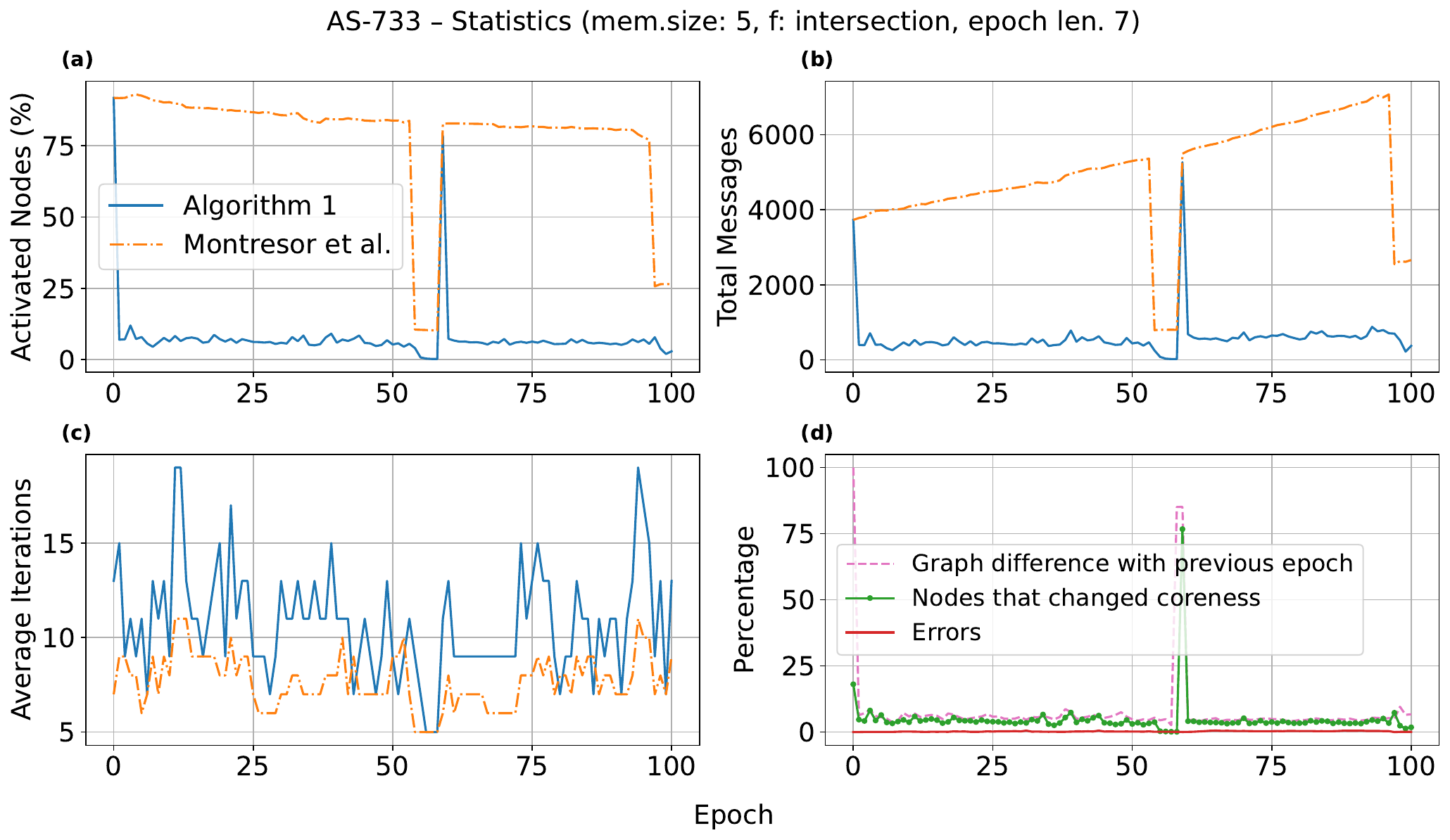}
    \caption{Visualization of the results obtained on the \emph{AS-733} dataset, for the edge aggregation function \emph{intersection}.}
    \label{fig:as733-comparison}
\end{figure}

Another thing we notice is the average number of iterations performed by Algorithm~\ref{alg:rucci} against the competitor: this is due to our strategy of delaying the action of sending an update to the next iteration, to reduce the error rate.
Although this may appear as a step back with respect to the Montresor et al. strategy \cite{montresor_distributed_kcore_2013}, it is important to note that despite this, the number of activated nodes and the total number of messages exchanged do not increase. 
Thus, the savings in resources are still consistent even if the number of iterations to reach convergence increases.
Moreover, in our experiments, this number never exceeded twice the amount of the competitor. 
Fig.~\ref{fig:as733-comparison}(c) shows the values of the number of iterations for the AS-733 dataset, where we can further verify this statement.

We now turn to the error rate. 
Recall that we count one error for each node that, at the end of an execution of Algorithm~\ref{alg:rucci}, has a coreness value different from the one computed by our competitor algorithm. 
Table~\ref{tab:experimental_results} shows the average number of errors per epoch, both in absolute and relative terms, relative to the number of nodes in the graph. 
We consistently obtained less than $1\%$ of errors in every configuration for every dataset, proving we can achieve a significant speedup without losing too much accuracy. 
Additionally, in our experiments, we verified that all errors are always $\pm 1$ with respect to the correct value of coreness computed by our competitor.
It is interesting to see how the algorithm reacts to different epoch lengths when the aggregation function and memory size are fixed.
Indeed, there is basically no difference between the results for the Reddit dataset when the epoch length is enlarged to 14 days instead of 7.
On the other hand, if we change the function and keep the same epoch length, things change drastically (e.g. Reddit dataset with intersection).

\begin{figure}[tb]
    \centering
    \includegraphics[width=\textwidth]{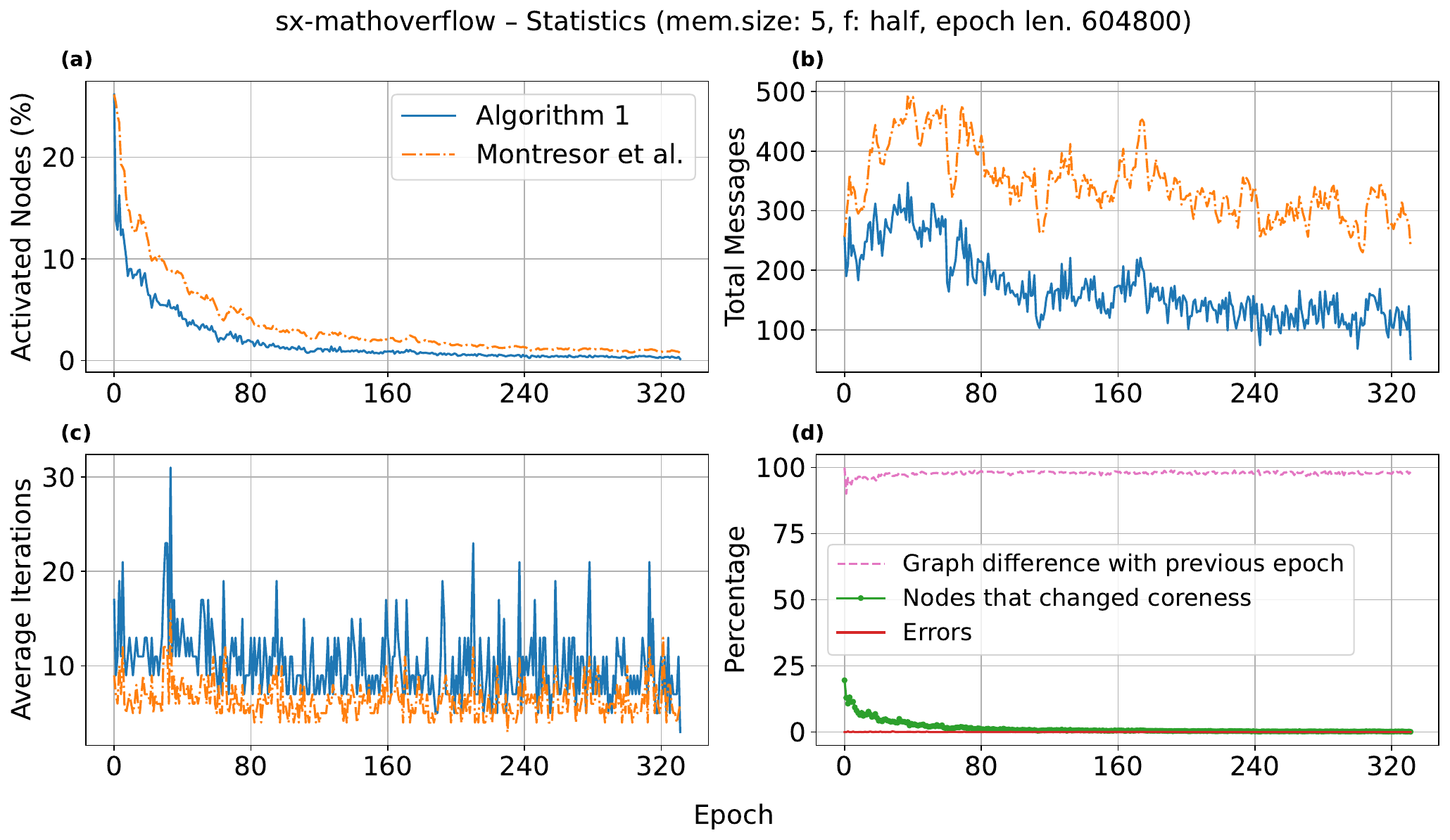}
    \caption{Visualization of the results obtained on the \emph{sx-mathoverflow} dataset, for the edge aggregation function \emph{union-2}}
    \label{fig:mathoverflow-comparison}
\end{figure}

Fig.~\ref{fig:as733-comparison} and Fig.~\ref{fig:mathoverflow-comparison} go more in depth on the statistics we gathered for two specific datasets and configurations, namely AS-733 with intersection as the edge aggregation function, and sx-mathoverflow with union.
In particular, we plotted the percentage of activated nodes for our algorithm and our competitor by Montresor et al. \cite{montresor_distributed_kcore_2013}, together with two measures of how much the corresponding graph has changed with respect to the previous epoch.
We first computed the Jaccard similarity between the edge list of any two consecutive epochs and subtracted it from 1 to obtain a measure of how much the graph has changed from the perspective of the edges. 
This number is trivially equal to 1 for the first epoch of every graph.

We also calculated the number of nodes that changed their coreness value with respect to the previous epoch: this number also includes those nodes who became isolated (i.e. have coreness equal to 0 or, equivalently, are not part of the graph anymore) on an epoch, hence the spikes in Fig.~\ref{fig:as733-comparison}. 
Correctly, the number of activated nodes stays low even in the presence of such spikes, because the isolated nodes do not send any message, therefore, they cannot become active.
On the other hand, we see that the trend of activated nodes in AS-733 (Fig.~\ref{fig:as733-comparison}) follows the trend of how much the edges in the graph change, as high values of dissimilarity (pink line) correspond to higher values of activated nodes by Algorithm~\ref{alg:rucci}. 
This holds for our competitor too, but that algorithm is blind to changes in the graph\footnote{For the AS-733 dataset, there is a significant change in the graph in epochs 54-59. This is intrinsic to the dataset itself and not caused by our implementation. The possible reasons behind this sudden change are discussed in \cite{Rucci:2024SAC}.}, as it recomputes everything from scratch each time.

\subsection{Takeaways}
The experimental results suggest two main conclusions about our approach:
\begin{enumerate}
\item Algorithm~\ref{alg:rucci} significantly reduces the number of exchanged messages compared to the competitor solution. This translates into lower computational demand and enables the analysis of larger graphs than previously feasible.

%
%
\item Although our algorithm requires slightly more iterations to converge, it maintains a low level of node activation throughout. As a result, the majority of nodes remain idle during execution, minimizing overall resource consumption and increasing scalability despite the increased number of rounds.
\end{enumerate}

        


\section{Conclusions}\label{sec:conclusions}
We presented a new decentralized algorithm for the maintenance of core decomposition in large temporal graphs, an important task for community detection and analysis in nowadays networks.
%
We conducted an extensive experimental phase, demonstrating significant performance improvements compared to the direct translation of an existing algorithm \cite{montresor_distributed_kcore_2013} into the temporal scenario. These improvements are measured in terms of the total messages exchanged during the algorithm's execution and the number of graph nodes that send at least one message.
While this strategy sometimes leads to small errors in the computed coreness values, we showed that these errors are \emph{(a)} always at $\pm 1$ unit from the true value, and \emph{(b)} extremely infrequent in our real-world dataset.
Our findings pave the way for future work on this algorithm, to ensure its correctness in all cases while maintaining the same desirable performance.

\bibliographystyle{abbrv}
\bibliography{ASONAM/lncs_biblio}

\end{document}